\documentclass[journal]{IEEEtran}
\IEEEoverridecommandlockouts

\usepackage{booktabs}
\usepackage{threeparttable}
\usepackage{amsmath,graphicx,cite,amssymb}
\usepackage{amsfonts,multirow,bm,array,setspace,stfloats}
\usepackage{graphicx,float,cite,amssymb}
\usepackage{graphics}
\DeclareGraphicsExtensions{.pdf,.jpeg,.png,.jpg}   
\usepackage{multirow,bm,bbm,array,setspace}
\usepackage{textcomp}
\usepackage{multirow}
\usepackage{diagbox}
\usepackage{psfrag}
\usepackage{color}
\usepackage{pstricks,enumerate}
\usepackage{bbm}
\usepackage{makecell}
\usepackage[normalem]{ulem}
\usepackage[caption=false,font=normalsize,labelfont=sf,textfont=sf]{subfig}
\usepackage{algorithm,algpseudocode}
\usepackage{float}
\usepackage{hyperref}
\usepackage{xpatch}
\makeatletter
\newcommand{\distas}[1]{\mathbin{\overset{#1}{\kern\z@\sim}}}%
\newsavebox{\mybox}\newsavebox{\mysim}
\newcommand{\distras}[1]{%
  \savebox{\mybox}{\hbox{\kern1pt$\scriptstyle#1$\kern1pt}}%
  \savebox{\mysim}{\hbox{$\sim$}}%
  \mathbin{\overset{#1}{\kern\z@\resizebox{\wd\mybox}{\ht\mysim}{$\sim$}}}%
}
\ExplSyntaxOn

\ExplSyntaxOn
\cs_new:Npn \bibColoredItems #1#2
  {
    \clist_map_inline:nn {#2} { \cs_new:cpn {bib@colored@##1} {#1} } 
  }
\ExplSyntaxOff

\newcommand\bib@setcolor[1]{%
  \ifcsname bib@colored@#1\endcsname
    \expandafter\color\expandafter{\csname bib@colored@#1\endcsname}
  \else
    \normalcolor
  \fi
}

\makeatother
\ifCLASSINFOpdf
\else
\fi

\newcommand{\bp}{\bm{p}}

\newcommand{\bu}{\bm u}

\newcommand{\by}{\bm{y}}

\newcommand{\bx}{\bm{x}}
\newcommand{\bA}{\bm{A}}

\newcommand{\bB}{\bm{B}}

\newcommand{\bbf}{\bm{f}}

\newcommand{\be}{\bm e}
\newcommand{\br}{\bm r}

\newtheorem{remark}{Remark}

\begin{document}

\title{A Semantic Approach to Successive Interference Cancellation for Multiple Access Networks}
\author{
 \IEEEauthorblockN{
 Mingxiao Li, \IEEEmembership{Graduate Student Member,~IEEE}, Kaiming Shen, \IEEEmembership{Senior Member,~IEEE},\\ and Shuguang Cui, \IEEEmembership{Fellow,~IEEE}
 } 
 
 \thanks{
    Manuscript published in IEEE Internet of Things Journal. This work was supported in part by NSFC under Grant 62293482 and Grant 12426306, in part by the Basic Research Project No. HZQB-KCZYZ-2021067 of Hetao Shenzhen-HK S\&T Cooperation Zone, and in part by the Guangdong Provincial Key Laboratory of Future Networks of Intelligence under Grant 2022B1212010001. Source code available at \href{https://github.com/lmx666-gif/semantic-SIC}{https://github.com/lmx666-gif/semantic-SIC}. \emph{(Corresponding author: Kaiming Shen.)}
            
    The authors are with the Future Network of Intelligence Institute (FNii), the School of Science and Engineering, The Chinese University of Hong Kong, Shenzhen, China (e-mail: mingxiaoli@link.cuhk.edu.cn; shenkaiming@cuhk.edu.cn; shuguangcui@cuhk.edu.cn). 
            }
 }


\maketitle

\begin{abstract}
Differing from the conventional communication system paradigm that models information source as a sequence of (i.i.d. or stationary) random variables, the semantic approach aims at extracting and sending the high-level features of the content deeply contained in the source, thereby breaking the performance limits from the  statistical information theory. As a pioneering work in this area, the deep learning-enabled semantic communication (DeepSC) constitutes a novel algorithmic framework based on the transformer---which is a deep learning tool widely used to process text numerically. The main goal of this work is to extend the DeepSC approach from the point-to-point link to the multi-user multiple access channel (MAC). The inter-user interference has long been identified as the bottleneck of the MAC. In the classic information theory, the successive interference cancellation (SIC) scheme is a common way to mitigate interference and achieve the channel capacity. Our main contribution is to incorporate the SIC scheme into the DeepSC. As opposed to the traditional SIC that removes interference in the digital symbol domain, the proposed semantic SIC works in the domain of the semantic word embedding vectors. Furthermore, to enhance the training efficiency, we propose a pretraining scheme and a partial retraining scheme that quickly adjust the neural network parameters when new users are added to the MAC. We also modify the existing loss function to facilitate training. Finally, we present numerical experiments to demonstrate the advantage of the proposed semantic approach as compared to the existing benchmark methods.
\end{abstract}


\section{Introduction}
\IEEEPARstart{I}{n} the past development of wireless communications, the classic information theory \cite{shannon1948mathematical} has played a central role in characterizing the fundamental limits for data transmission. The classic information theory heavily relies on the following assumption: the information source amounts to an i.i.d. or stationary stochastic process. However, the real-world communication systems are much more complicated than the above ideal model. Unlike the traditional bit-based communication schemes, semantic communication  \cite{tong2022nine,lu2023semantics} aims to transmit the semantic meaning embedded in the information source, not requiring any specific statistical model for the information source. This paper proposes a semantic extension of the successive interference cancellation (SIC) scheme for the multiple access channel (MAC) with multiple users.

A comprehensive portrayal of semantic communications in the 6G systems can be found in \cite{wang2023road}. For instance, \cite{weng2021semantic} proposes a semantic communication system using an attention mechanism to reconstruct signals, \cite{luo2022autoencoder} considers an auto-encoder neural network for semantic relaying, \cite{basu2014preserving} proposes using the so-called semantic entropy to quantify the limit of semantic compression, \cite{chaccour2024less} devises a reasoning approach to semantic communications, and \cite{yang2023energy} views the energy efficiency problem from a semantic perspective. Moreover, the application of deep reinforcement learning in task-oriented semantic communication is considered in \cite{zhang2023drl}, and the multi-modal algorithms for bidirectional caching are examined \cite{wang2023multimodal}. It is also shown that semantic embedding leads to superior performance in source data compression \cite{crossword} and significantly reduces the required communication resources \cite{semantic_survey}. Further, deep learning has been used as a major tool in a variety of semantic communication tasks, such as the text \cite{xie2021deep}, images \cite{hu2022robust}, speech \cite{weng2021semantic}, and video \cite{niukai_video}. Semantic communications are also discussed in more sophisticated application scenarios, such as the digital twins \cite{du2023yolo,chen2023networking}, the metaverse \cite{ali2023metaverse,wang2023semantic}, wireless sensing \cite{du2023semantic}, and security transmission \cite{chen2023deep}.

Aside from the above works on the link-level semantic communication, some other recent works concern the multi-link case. For instance, \cite{xie2022task} explores task-oriented multi-user semantic communications to transmit data with single-modality and multiple modalities. In \cite{lin2023channel}, a semantic framework is proposed to eliminate the excessive training on different channels for the multi-user orthogonal frequency-division multiplexing-nonorthogonal multiple access (OFDM-NOMA) systems. A so-called semantic feature division multiple access (SFDMA) scheme is proposed in \cite{ma2024semantic} to allow multiple users to share the same time-frequency resource, while the multi-user efficient transmission
DeepSC scheme in \cite{huang2024flag} works for multiple semantic transceivers of separate private messages. Moreover, \cite{mu2023exploiting} proposes a semantic implementation of the authorized uplink NOMA system. For the broadcast channel, \cite{wu2023fusion} proposes an image semantic fusion scheme. Based on the vision transformer (ViT), \cite{wu2024deep} devises a self-attention mechanism for the multi-user multiple-input multi-output (MIMO) channels.

\begin{table*}[htbp]\renewcommand\arraystretch{1.2}
\small
\centering
\caption{List of Main Variables}
\begin{tabular}{|c||l|}
\hline
\textbf{Symbol} & \textbf{Definition} \\ \hline
\hline
$K$ & number of user terminals \\ \hline
$h_i$ & the channel from $i$th user terminal to BS \\ \hline
$M$ & number of channel uses \\ \hline
$T_i (\widehat{T}_i)$ & transmitted (received) text of $i$th user terminal \\ \hline
$N_i$ & number of words in the text of $i$th user terminal \\ \hline
$y_m$ & received uplink signal at BS for $m$th channel use \\ \hline
$P_i$ & power constraint of $i$th user terminal \\ \hline
$\be^{i}_{j,\ell} (\widehat{\be}^{i}_{j,\ell})$ & one-hot embedding vector of $\ell$th word of $j$th sentence from $i$th user terminal \\ \hline
$\bm{f}^i_{j,\ell} (\widehat{\bm{f}}^i_{j,\ell})$ & word embedding of the $\ell$th word of the $j$th sentence from the $i$th user terminal \\ \hline
$\bm{u}^i_{j,\ell} (\widehat{\bm{u}}^i_{j,\ell})$ & word semantic vector of $\ell$th word of $j$th sentence from $i$th user terminal \\ \hline
$\bm{r}^i_{j,\ell} (\widehat{\bm{r}}^i_{j,\ell})$ & latent variable that acts as semantic compression of $\bm{u}^i_{j,\ell}$ \\ \hline
$\bm{q}^i_{j} (\widehat{\bm{q}}^i_{j})$ & long vector that formed by $\bm{r}^i_{j,\ell}$ in $j$th sentence from $i$th user terminal \\ \hline
$\bm{x}^i_{m}$ & $m$th complex-valued symbol from $T_i$ \\ \hline
$\widetilde{\bm r}_j$ & feature served as side information for $j$th sentence \\ \hline
$\mathrm{TE}_i(\cdot)$ & semantic encoder of $i$th user terminal \\ \hline
$\mathrm{AE}_i(\cdot)$ & channel encoder of $i$th user terminal \\ \hline
$\mathrm{AD}_i(\cdot)$ & channel decoder of $i$th user terminal \\ \hline
$\mathrm{TD}_i(\cdot)$ & semantic decoder of $i$th user terminal \\ \hline
$\mathrm{IFG}_i(\cdot)$ & integrated feature generator of $i$th user terminal \\ \hline
$\Omega_i(\cdot)$ & neural network that extracts feature information from recovered text in $\mathrm{IFG}_i$ \\ \hline
$\theta_i(\cdot)$ & neural network that extracts feature information from a concatenation of outputs of $\Omega$ and $\pi_i$ in $\mathrm{IFG}_i$ \\ \hline
$\pi_i(\cdot)$ & neural network that extracts feature information from current text in $\mathrm{IFG}_i$ \\ \hline
$\bm g^i_{j}$ & feature generated by $\mathrm{IFG}_i$ to facilitate decoding the $j$th sentence of user terminal $i$ \\ \hline
\end{tabular}
\end{table*}

The semantic method proposed in this paper is based on the Transformer networks \cite{vaswani2017attention}---which is already extensively applied to the communication system design. For example, \cite{gu2023spatial,zhou2022fedformer} suggest using the Transformer to estimate channels. The authors of \cite{wu2024transformer} propose a deep learning paradigm for the task of image transmission with feedback. For the reconfigurable intelligent surface (RIS)-assisted wireless network, \cite{xie2022quan} proposes a quantized Transformer method for channel estimation, and \cite{peng2022robust} devises a Transformer-based error-correcting decoder. In particular, the Transformer has been shown fairly useful in the deep learning-enabled semantic communication (DeepSC) framework \cite{xie2021deep} for extracting the semantic feature from the raw data source. Moreover, a universal Transformer is used to extract semantic features in \cite{xie2023semantic}, while a lite distributed implementation of the DeepSC is proposed in \cite{xie2020lite}. Our work is most closely related to the DeepSC \cite{xie2021deep}. As opposed to the DeepSC that focuses on the point-to-point transmission, this work extends the Transformer based semantic extraction for the multi-user communications in the MAC channels, with the SIC scheme incorporated in. We focus on the MAC because SIC is proved to be capacity achievable for this particular type of channel. Regarding other types of multi-user channels, such as interference and broadcast network, the proposed method can also be applied just as how SIC is applied to them, only that even the SIC may not guarantee the achievability of channel capacity. In principle, our work differs from the previous semantic communications methods in the following three respects:
\begin{itemize}
    \item \emph{Generic Multiple Access Channel:} The previous work \cite{mu2023exploiting} considers a two-user multiple access channel in which only one user follows the semantic scheme while the other still uses the traditional scheme. In contrast, this work considers a general $K$-user multiple access channel, and assumes that all the transmissions are semantic.
    \item \emph{Semantic Successive Interference Cancellation:} The previous work \cite{liang2023semantic} integrates the SIC technology into the semantic scheme for MIMO transmission. However, the SIC in \cite{liang2023semantic} is the traditional one in that it simply models the different information sources as i.i.d. random variables, whereas our work proposes a novel semantic SIC scheme that captures the semantic correlation between the information sources.
    \item \emph{Dynamic System Setting:} The initial end-of-end semantic communication paradigm in \cite{xie2021deep} has been extended for a variety of multi-user systems, e.g., the two-user multiple access \cite{mu2023exploiting}, and the MIMO spatial multiplexing \cite{xie2022task}, \cite{liang2023semantic}. However, to the best of our knowledge, all these extensions assume fixed system setting. Thus, when a new user joins the system, the existing methods require retraining the whole system from scratch. In contrast, this work proposes an efficient adaptation without retraining the encoders and decoders.
\end{itemize}

The main contributions of this work are summarized in the following:
\begin{itemize}
\item We extend the DeepSC scheme (which is originally designed for an isolated point-to-point link) to the MAC case. In particular, we propose performing the SIC in the semantic domain (more specifically, in terms of the semantic work embedding vector) to mitigate the inter-link interference.
\item By the conventional SIC scheme, once the message of a particular user has been recovered, we just remove its corresponding signal from the received signal and then do not use it anymore in the message decoding for other users. While doing so is optimal when the multiple sources are modeled as independent random variables, we point out that the already decoded message could facilitate the later decoding by leveraging the semantic correlation between the sources, and thereby propose a side-link assisted semantic decoding scheme.
\item When extending the DeepSC scheme for multiple links, a natural idea is to train the neural networks for all links fully. However,
the adaptability of this full training strategy poses a big challenge, i.e., we must retrain everything whenever a new user is added to the MAC. To address this issue, we propose a partial retraining scheme. Furthermore, we devise a novel loss function tailored to the retraining task.
\end{itemize}

The rest of the paper is organized as follows. Section \ref{sec: framework} describes the system model of the MAC from a semantic perspective, in particular with the semantic performance metric specified. Section \ref{sec:encoder} focuses on the encoding part of the semantic MAC system, while Section \ref{sec:decoder} focuses on the decoding part. Section \ref{sec:training} discusses how to train the neural networks across multiple links. Section \ref{sec:experiments} presents the experiment results. Finally, Section \ref{sec:conclusion} concludes this work.

\begin{figure*}[t]
\centering
\hspace{-0.3cm}

\includegraphics[width=0.8\linewidth]{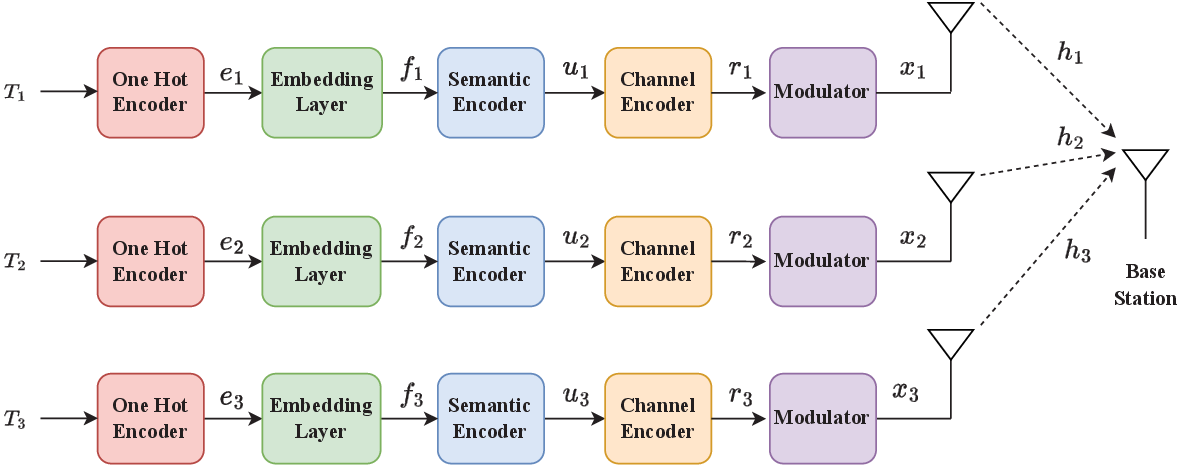}

\caption{Semantic communications in a 3-user MAC.}
\label{fig:system_encoder}
\end{figure*}

\section{System Model}
\label{sec: framework}

Consider an uplink network wherein the base station (BS) is associated with $K$ user terminals. Denote by $h_i\in\mathbb C$, $i=1,2,\ldots,K$, the channel from the $i$th user terminal to the BS. 
Assume without loss of generality that the user terminals satisfy:
\begin{equation}
\label{sic_seq}
    P_1|h_1|^2 \ge P_2|h_2|^2\ge \ldots \ge P_K|h_K|^2.
\end{equation}
Clearly, we can always render the user terminals satisfy the above inequality chain by re-indexing them properly. 
Assume that these user terminals wish to send private (English) texts toward the BS simultaneously at the same spectrum. The text of the $i$th user terminal is denoted by $T_i$, which comprises $N_i$ words. In order to send the text over wireless channel, the $i$th user terminal converts its text $T_i$ to a sequence of $M$ complex-valued symbols $(x^i_1,x^i_2,\ldots,x^i_M)^\top$:
\begin{equation}
     \mathrm{ENC}_i(T_i)=\begin{bmatrix}
        x^i_1\\ x^i_2\\ \vdots\\ x^i_M
    \end{bmatrix}\in\mathbb C^{M},
\end{equation}
which can be thought of a joint source-and-channel encoder. A power constraint $P_i$ is imposed on each user terminal $i$, i.e.,
\begin{equation}
    \frac{1}{M}\sum^M_{m=1} |x^i_m|^2 = P_i.
\end{equation}
Thus, for the $m$th channel use, the received uplink signal at the BS is given by:
\begin{equation}
    y_m = \sum^K_{i=1}h_ix^i_m+z_m,
\end{equation}
where $z_m\sim\mathcal{CN}(0,\sigma^2)$ is an i.i.d. Gaussian noise across the $M$ channel uses, and $\sigma^2$ is the noise power.

The BS aims to recover all the texts $\{T_i\}$ from the received signal vector $(y_1,y_2,\ldots,y_M)^\top$:
\begin{equation}
     \mathrm{DEC}(\begin{bmatrix}
        y_1\\ y_2\\ \vdots\\ y_M
    \end{bmatrix}) = \{\widehat{T}_1,\widehat{T}_2,\ldots,\widehat{T}_K\},
\end{equation}
where $\widehat{T}_i$ denotes the decoded text of $T_i$.

We consider the optimal design of $\textsf{ENC}_i(\cdot)$ and $\textsf{DEC}(\cdot)$ given the fixed $K$, $\{h_i\}$, $M$, $\sigma^2$, and $\{P_i\}$. A traditional information-theoretic setup is to:
\begin{subequations}
\begin{align}
\underset{}{\text{find}}&\quad   \mathrm{ENC}_1(\cdot),\ldots, \mathrm{ENC}_K(\cdot), \mathrm{DEC}(\cdot)\\
\text{subject to}&\quad \mathrm{Pr}\{T_i\ne\widehat{T}_i\}\le \epsilon,\;\text{for } i = 1,2,\ldots,K,
\end{align}
\end{subequations}
where $\epsilon>0$ is the given outage probability target.

In contrast, this work adopts a semantic objective. Following the previous work \cite{xie2021deep}, we design $ \mathrm{ENC}_i(\cdot)$ and $ \mathrm{DEC}(\cdot)$ to maximize the \emph{semantic similarity} between $T_i$ and $\widehat{T}_i$ for each user terminal $i$; an existing definition of the semantic similarity is stated as follows. First, we parse the text into 
sentences, i.e.,
\begin{align}
    T_i &= \left\{S^i_1, S^i_2\ldots,S^i_{L_i}\right\},\\
    \widehat{T}_i &= \left\{\widehat{S}^i_1, \widehat{S}^i_2\ldots,\widehat{S}^i_{L_i}\right\},
\end{align} 
where $L_i$ is the number of sentences contained in the text. Then for each sentence $S^i_j$ or $\widehat{S}^i_j$, we use the \emph{bidirectional encoder representation (BERT)} \cite{reimers2019sentence} to map it to a $384\times1$ real-valued vector, denoted by $\mu^i_j\in\mathbb R^{384\times1}$ or $\widehat\mu^i_j\in\mathbb R^{384\times1}$. Next, we compute the \emph{cosine similarity} \cite{rahutomo2012semantic} (a.k.a. the BERT similarity) between $\mu^i_j$ and $\widehat{\mu}^i_j$ as:
\begin{equation}
    \lambda(S^i_j,\widehat{S}^i_j) = \frac{({\bm\mu_j^i})^\top\widehat{\bm\mu}^i_j}{\|\bm\mu^i_j\|_2 \times\|\widehat{\bm\mu}^i_j\|_2}.
\end{equation}
Observe that $0\le \lambda(S^i_j,\widehat{S}^i_j) \le 1 $. Note that the BERT similarity is the state-of-the-art semantic measure for the text message, which is extensively used in the literature on semantic communications \cite{xie2021deep,xie2022task}. In the paper, we also use the bilingual evaluation understudy (BLEU) score \cite{bleu} to measure the semantic similarity. BERT \cite{reimers2019sentence} and BLEU are more suited than other similarity metrics (e.g., CIDEr \cite{vedantam2015cider}) for semantic communications because they effectively capture the context relevance. In particular, BERT is also widely adopted in the large language model (LLM) \cite{achiam2023gpt} area for generating human-like contents because of its strong capability to quantify the inherent connections between context and each single word. Nevertheless, in the revised manuscript, we also clarify that BERT and BLEU are limited to the text data; if other modes of messages are considered, we need to change the similarity metric accordingly, e.g., SDR \cite{weng2021semantic} for audio and SSIM \cite{yang2021deep} for image. The semantic similarity is defined to be the average cosine similarity across the sentences within the text, i.e., 
\begin{equation}
\label{simi}
    \delta(T_i,\widehat T_i) = \frac{1}{L_i}\sum^{L_i}_{j=1}\lambda(S^i_j,\widehat{S}^i_j).
\end{equation}
The aim of this work is to seek the optimal encoder-and-decoder pair under the semantic similarity constraint $0\le \zeta\le 1$ across the $K$ user terminals, i.e., $\min_i\left\{\delta(T_i,\widehat T_i)\right\}\ge \zeta$.

\begin{remark}[Connection to LLM]
    The proposed semantic communication method has intimate connections with the LLM because they both employ the Transformer of the BERT to convert each word to a ``word semantic vector'' in accordance with the specific context. Nevertheless, our method and the LLM differ in two respects. First, their model sizes differ dramatically. Our method is aimed at the interference cancellation for a particular communication system of multiple access channel, for which the 4-layer Transformer is sufficient as shown in the paper. In contrast, the task of the LLM is much more generic and requires a much more sophisticated neural network, e.g., GPT-4 uses up to 120 Transformer layers \cite{achiam2023gpt}. Second, our method and the LLM utilize the word semantic vector for distinct purposes. Our method treats the word semantic vector as a sequence of symbols transmittable over wireless channels, thus eliminating the conventional model-based signal modulation. In contrast, the LLM takes the word semantic vector as the input for the generative model to create human-like text and contents. Clearly, these generative contents heavily depend on the context and thus the BERT plays an important role. A new insight proposed in this paper is that the interference cancellation is closely related to the context too. The above points are added to the new manuscript.
\end{remark}

\begin{figure*}[t]
\centering
\includegraphics[width=1.0\linewidth]{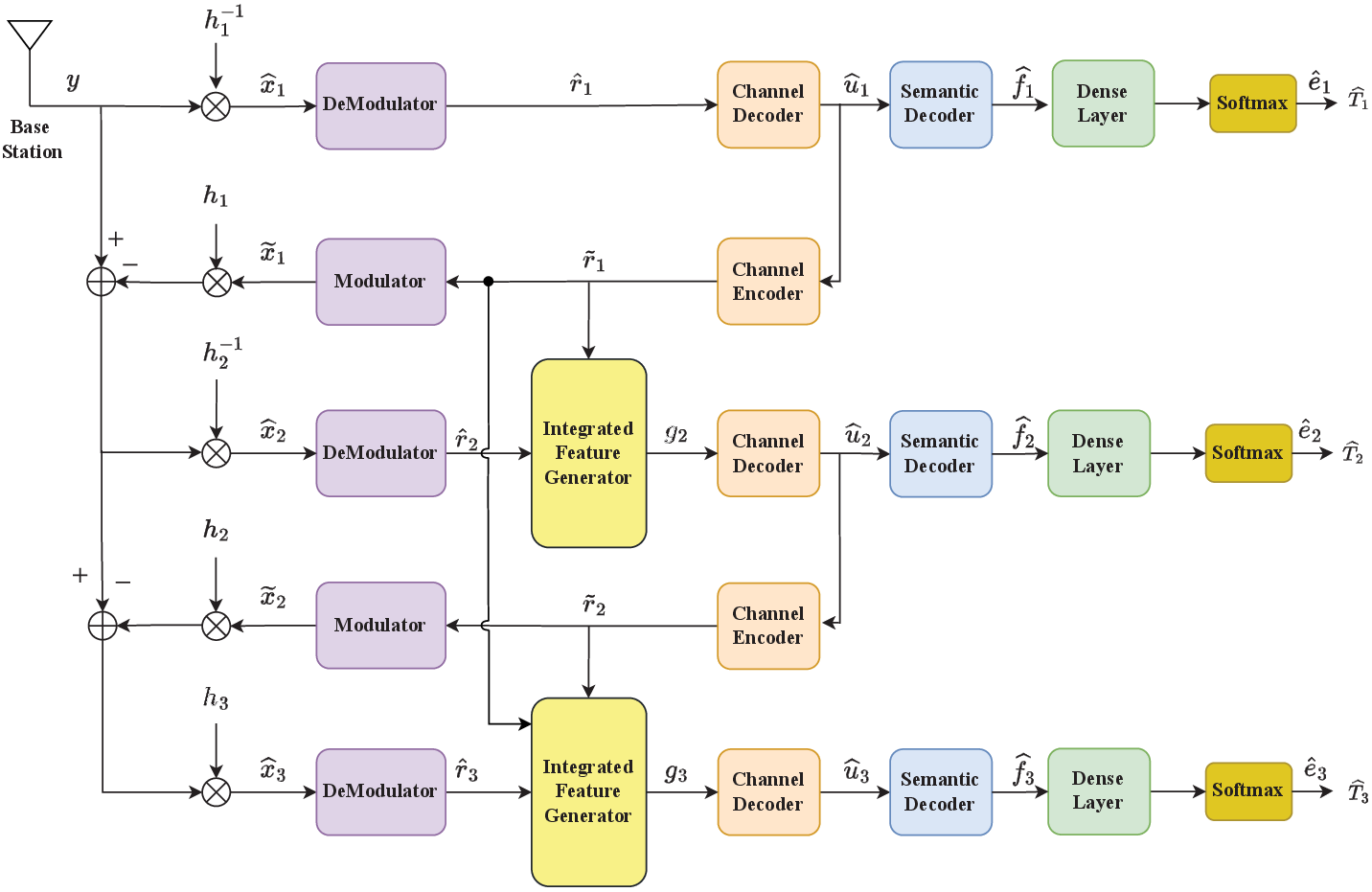}
\caption{The paradigm of the proposed semantic decoder for the 3-user MAC.}
\label{fig:system_decoder}
\end{figure*}

\section{Semantic Encoding}
\label{sec:encoder}

We begin with the encoding part. As shown in Fig.~\ref{fig:system_encoder}, this example illustrates the scenario for three users. Let $\mathcal W$ be an ordered set of all the words that could possibly appear in the texts, i.e.,
\begin{equation}
    \mathcal{W} = \{V_0,V_1,V_2,\ldots,V_{|\mathcal W|-1}\},
\end{equation}
which is also called the dictionary. In particular, $V_0$ is preserved for the special symbol ``$\langle$END$\rangle$'' that marks the end of each sentence. Each word $V_i$ can then be represented by its position in the dictionary; we propose converting each word to a $|\mathcal W|$-dim vector whose $i$th entry equals one while the rest entries all equal zero, which is known as the \emph{one-hot embedding vector} \cite{peng2022robust}. We denote by $\be^{i}_{j,\ell}\in\{0,1\}^{|\mathcal W|}$ the one-hot embedding vector of the $\ell$th word of the $j$th sentence from the $i$th user terminal.

We now have a numerical representation for the text, but the corresponding one-hot embedding vector is difficult to tackle because it is quite lengthy (whose length equals the size of the dictionary, $|\mathcal W|$). We now seek a more efficient numerical representation of the text by exploring the word structure, namely the \emph{word embedding}. First, we compress the one-hot embedding vector as:
\begin{equation}
\label{f=Ae}
    {\widetilde\be}^i_{j,\ell} = {\bm A}_i \be^{i}_{j,\ell}\in\mathbb R^{d},
\end{equation}
where $\bm A_i\in\mathbb R^{d\times|\mathcal W|}$ is a matrix to be optimized by the neural network, and $d$ is a given integer and is typically much smaller\footnote{In our case, $|\mathcal W|\approx23,000$ with $d=128$ by default.} than $|\mathcal W|$; we remark that $\bA_i$ is user terminal specific since it depends on the context of each $T_i$. The optimization of $\bm A_i$ is discussed in detail in Section~\ref{sec:training}.

The second stage of word embedding is to perform \emph{position encoding} (PE) \cite{vaswani2017attention} for the Transformer encoding purpose. Specifically, for the $\ell$th word of each sentence, we generate a PE sequence $\bp_{\ell}\in\mathbb R^{d}$ as follows:
\begin{equation}
    \bp_\ell = (p_{\ell,1},p_{\ell,2},\ldots,p_{\ell,d})^\top,
\end{equation}
where
\begin{equation}
p_{\ell,z} = \begin{cases}
      \sin(\ell\times10^{-{4z}/{d}}) & \text{if $z\equiv0\mod 2$}\\
      \cos(\ell\times10^{-{4(z-1)}/{d}}) & \text{if $z\equiv1\mod 2$}
    \end{cases},
\end{equation}
for $z=1,2,\ldots,d$. For the $\ell$th word of the $j$th sentence of $T_i$, we obtain its word embedding vector as: 
\begin{equation}
    \bbf^i_{j,\ell} = {\widetilde\be}^i_{j,\ell} + \bp_{\ell}.
\end{equation}
In summary, ${\widetilde\be}^i_{j,\ell}$ can be thought of as a numerical label of the present word, while $\bp_{\ell}$ can be thought of the position of this word in the sentence it belongs to, so $\bbf^i_{j,\ell}$ contains both the word information and the position information.

\begin{figure*}[t]
\includegraphics[width=1.0\linewidth]{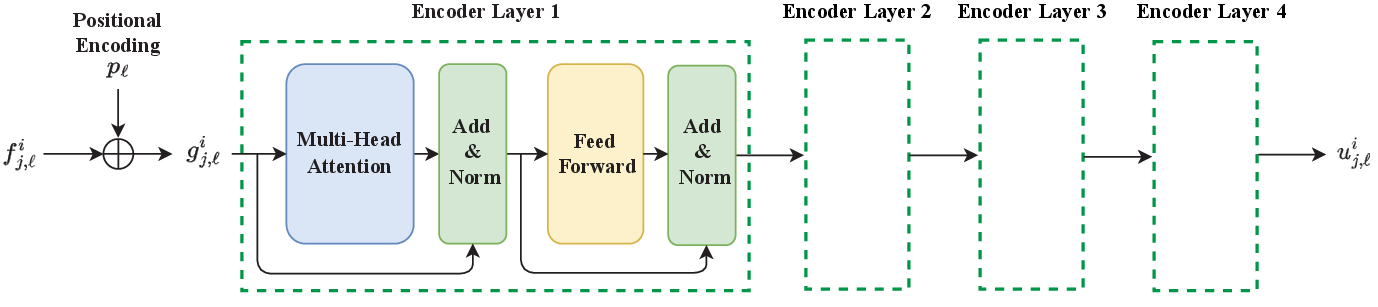}
\caption{The paradigm of the proposed semantic encoder. The dashed boxes represent the replicas of the first dashed box whose inside structure is depicted.}
\label{fig:semantic_encoder}
\end{figure*}

The next task is to extract the semantic feature (or roughly speaking, the meaning) of the present word from its embedding vector $\bbf^i_{j,\ell}$. The Transformer encoder \cite{vaswani2017attention} is employed to achieve this. Just like the one-hot embedding vector compressor $\bA_i$ in \eqref{f=Ae}, the Transformer encoder differs from user terminal to user terminal because it depends on the context of $T_i$ of each user terminal. Denoting the Transformer encoder assigned to user terminal $i$ by $ \mathrm{TE}_i$, depicted in Fig.~\ref{fig:semantic_encoder}, we take the word embedding vector $\bbf^i_{j,\ell}\in\mathbb R^d$ as input and then obtain the \emph{word semantic vector} $\bu^i_{j,\ell}\in\mathbb R^m$:
\begin{equation}
    \bu^i_{j,\ell} =  \mathrm{TE}_i(\bbf^i_{j,\ell}),
\end{equation}
where $m$ is a given positive integer; a common choice is to let $m=d$.

\begin{figure*}[t]
\centering
\includegraphics[width=1.0\linewidth]{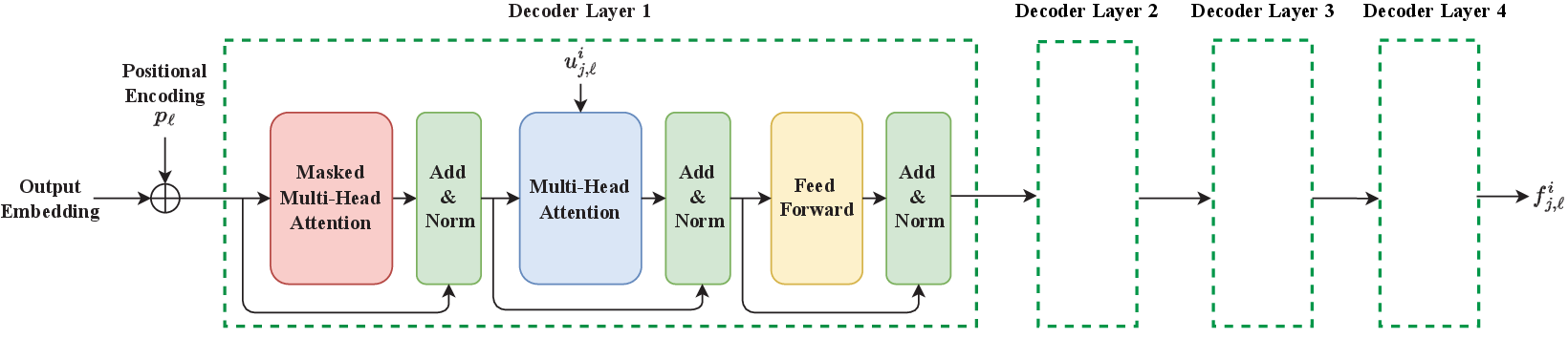}
\caption{The structure of semantic decoder. The dashed boxes represent the replicas of the first dashed box whose inside structure is depicted.}
\label{fig:semantic_decoder}
\end{figure*}

We wish to further compress the semantic representation of each word in order to enhance the transmission efficiency over wireless channels. The main idea is to apply the autoencoder. For some even integer $c<m$, let $ \mathrm{AE}_i:\mathbb R^m\rightarrow \mathbb R^c$ be the encoder component of the autoencoder for $T_i$, and let $ \mathrm{AD}_i:\mathbb R^c\rightarrow \mathbb R^m$ be the decoder component of the autoencoder for $T_i$. The autoencoder is then structured as:
\begin{align}
{\bm r}^i_{j,\ell} &=  \mathrm{AE}_i(\bm u^i_{j,\ell}),
    \label{AE}\\
\widehat{\bm u}^i_{j,\ell} &=  \mathrm{AD}_i(\bm r^i_{j,\ell}),
    \label{DE}
\end{align}
where the latent variable ${\bm r}^i_{j,\ell}$ acts as the semantic compression of $\bm u^i_{j,\ell}$. The bottomline is to train the neural networks of $ \mathrm{AE}_i$ and $ \mathrm{AD}_i$ together to minimize the distortion between $\bm u^i_{j,\ell}$ and $\widehat{\bm u}^i_{j,\ell}$. The trained $ \mathrm{AE}_i$ is then deployed at user terminal $i$, used to compress the corresponding word semantic vector $\bm u^i_{j,\ell}$.

After obtaining the compressed word semantic vector ${\bm r}^i_{j,\ell}\in\mathbb R^c$, we are ready to modulate it into complex symbols transmittable over wireless channels. Recall that each sentence $S^i_j$ consists of $N^i_j$ words. We recast each word of $S^i_j$ to $\bm r^i_{j,\ell}\in\mathbb R^c$ as discussed above, and then concatenate these word semantic vectors together to form a long vector $\bm q^i_j\in\mathbb R^{cN^i_j}$. In order to notify the receiver where the current sentence ends (in case that the special symbol $\langle$END$\rangle$ is missed), we perform zero padding to make the vector length equal to $cN$:
\begin{equation}
    \bm q^i_j = \begin{bmatrix}
        \bm q^i_j\\
        0\\
        \vdots\\
        0
    \end{bmatrix}\in\mathbb R^{cN}.
\end{equation}
In other words, we append $N-N^i_j$ dummy words to $S^i_j$ if $N^i_j<N$, and then recast each dummy word to a zero vector. Next, we sequentially take two entries of $S^i_j$, treating them as the real part and the complex part of a complex symbol. Thus, each sentence $S^i_j$ is converted to $Nc/2$ complex symbols $(\check x^i_{j,1},\check x^i_{j,2},\ldots,\check x^i_{j,\frac{Nc}{2}})$, with each $\check x^i_{j,t}$ given by:
\begin{equation}
\check x^i_{j,t} = \bm q^i_j[2t-1] + \mathrm{j}\cdot\bm q^i_j[2t],
\end{equation}
where $\bm q^i_j[b]$ is the $b$th entry of $\bm q^i_j$, and $\mathrm{j}$ is the imaginary unit. At each user terminal $i$, streaming out the symbol sequence $(\check x^i_{j,1},\check x^i_{j,2},\ldots,\check x^i_{j,\frac{Nc}{2}})$ across all the sentences $S^i_j$ within $T_i$ yields:
\begin{equation}
    \check \bx_i = (\check x^i_{1,1},\ldots,\check x^i_{1,\frac{Nc}{2}},\check x^i_{2,1},,\ldots,\check x^i_{2,\frac{Nc}{2}},\ldots)^\top.
\end{equation}
Ultimately, we obtain the transmit symbol sequence for user terminal $i$ by normalizing $\check \bx^i$ to the target power:
\begin{equation}
    (x^i_1,x^i_2,\ldots,x^i_M) = \frac{\sqrt{MP_i}}{\|\check \bx_i\|}\cdot\check \bx_i,
\end{equation}
where for convenient, we assume all $L_i$ are equal to $L$, so that $M=cNL/2$.

\section{Semantic Decoding}
\label{sec:decoder}

The main idea of the traditional SIC follows: first decode $T_1$, then subtract its signal from the received signal, and further decode $T_2$, and so forth. Fig.~\ref{fig:system_decoder} illustrates the decoding procedure for a three-user system. A major contribution of this work is to show that the SIC scheme can be considered in the semantic domain.

\subsection{Reversing the Semantic Encoding Procedure}
\label{subsec:reverse}

We begin with the decoding of $T_1$. Following the traditional SIC case, we just recover $T_1$ directly from the received signal, without considering other texts. Nevertheless, we adopt a semantic decoder here, rather than the traditional probability-based decoders (such as the LDPC or the typicality decoder from the information theory). 

The bottomline is fairly simple: we just reverse the operations in semantic encoding part. Recall the procedure of semantic encoding:

\begin{equation}
\label{se}
    T_i \rightarrow \be_i \rightarrow \bm f_i \rightarrow \bu_i
    \rightarrow \bm r_i \rightarrow \bm q_i\rightarrow \bx_i.
\end{equation}

Now let us see how user terminal 1 reverses the semantic encoding procedure and thereby recovers $T_1$. It first estimates $\bx_1$ as:
\begin{equation}
    \widehat \bx_1 = \frac{\by}{h_1}.
\end{equation}
Clearly, $\widehat \bx_1$ is exactly $\bx_1$ when the rest user terminals' transmitted symbols and the additive noise are negligible. By separating the real part and the imaginary part of $\widehat \bx_1$, we readily obtain an estimate of a long sequence of compressed word semantic vectors of sentences, written as $\widehat{\bm q}_1\in\mathbb R^{2M}$. We then reshape $\widehat{\bm{q}}_1$ into vectors of $\mathbb{R}^{L \times N \times c}$ because each sentence has been padded into $N$ words, and each word is represented by a vector $\widehat{\bm{r}}^1_{j,l} \in \mathbb{R}^c$, identical to $\bm{r}^1_{j,l}$.


Next, the decoder component of the autoencoder in \eqref{DE} is used to decompress $\widehat{\bm r}^1_j$ into $\widehat{\bm u}^1_j$---which mimics the word semantic vector:
\begin{equation}
    \widehat{\bm u}^1_{j,\ell} =  \mathrm{AD}_1(\widehat\br^1_{j,\ell}).
\end{equation}
Moreover, by the Transformer decoder \cite{vaswani2017attention} as shown
in Fig.~\ref{fig:semantic_decoder}, we recover $\widehat{\bm f}^i_{j,\ell}$
from $\widehat{\bm u}^i_{j,\ell}$:

\begin{equation}
\label{TD}
    \widehat{\bbf}^1_{j,\ell} =  \mathrm{TD}_1(\widehat{\bm u}^1_{j,\ell}).
\end{equation}

We now reverse the matrix multiplication in \eqref{f=Ae} as:
\begin{equation}
\label{softmax}
    \widehat \be^1_{j,\ell} = \sigma(\bB_1\widehat{\bm f}^1_{j,\ell}),
\end{equation}
where the softmax function $\sigma:\mathbb R^{|\mathcal W|}\rightarrow[0,1]^{|\mathcal W|}$ maps each entry of the vector to a value between 0 and 1, and $\bB_1\in\mathbb R^{|\mathcal W|\times d}$ can be thought of as a pseudo-inverse-like matrix of $\bA_1$ and shall be determined in conjunction with $\bA_1$ as specified later on. We remark that $\widehat \be^1_{j,\ell}\in\mathbb R^{|\mathcal W|}$ is a soft decision. To obtain a hard decision, we just pick the max entry and set it to 1, while letting the rest entries be zeros, namely the one-hot representation:
\begin{equation}
\label{sft_decision}
    \widehat \be^1_{j,\ell} = \text{one-hot}(\widehat \be^1_{j,\ell}).
\end{equation}
With the one-hot representation vectors in hand, we can readily recover $T_1$ based on the dictionary. In particular, whenever the special symbol $\langle \text{END} \rangle$ is detected in a sentence, any word following it will be removed.

\subsection{Semantic SIC with Assistance of Side Information}

\begin{figure*}[t]
\includegraphics[width=1.0\linewidth]{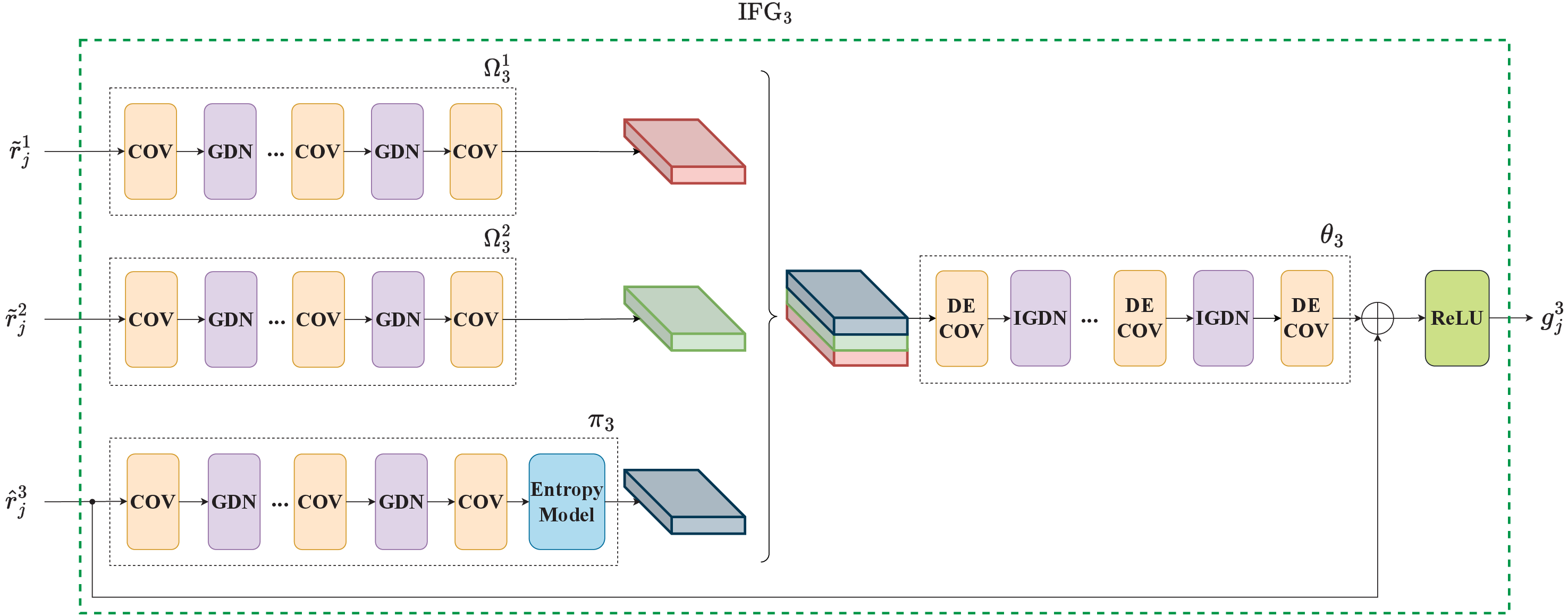}
\caption{Paradigm of the integrated feature generator $\mathcal{IFG}_3$ for user $3$, where COV denotes the convolution layer, GDN denotes the generalized divisive normalization layer, DECOV denotes the deconvolution layer, and IGDN denotes the inverse generalized divisive normalization layer \cite{balle2017end}.}
\label{fig:deepsc_si_combiner3}
\end{figure*}

After $T_1$ has been recovered, we then shift our attention to the second strongest user terminal and its corresponding text $T_2$. The key is how to take advantage of the already decoded $T_1$ to facilitate the decoding of $T_2$.

It is natural to perform the conventional symbol-level SIC, i.e., we redo the encoding process of $T_1$ and then remove the resulting complex symbols from the received signal $\bm y$. Thus, letting $\widetilde{\bm x}_1$ be the encoded symbols of $\widehat{T}_1$. We propose generating  $\widetilde{\bm x}_1$ as:
\begin{equation}
   \widetilde{\bm r}_1 =  \mathrm{AE}_1(\widehat{\bm u}_1).
\end{equation}
Following the subsequent procedure, $\widetilde{\bm x}_1$ can be generated from $\widetilde{\bm r}_1$ as outlined in \eqref{se}. We cancel the interference from $T_1$ as:
\begin{equation}
    \bm y = \bm y - h_1\widetilde{\bm x}_1,
\end{equation}
thus obtaining the estimate of $\bm x_2$:
\begin{equation}
    \widehat{\bm x}_2 = h_2^{-1}\bm{y}.
\end{equation}
We now take the above $\widehat{\bm x}_2$ as the input of the semantic decoder for $T_2$. The semantic decoder for $T_2$ has the same structure as that for $T_1$ as described in Section \ref{subsec:reverse}. The above symbol-level SIC can be readily extended for the subsequent user terminals. Next, for user terminal 3, we re-perform the encoding process on the recovered $T_2$ to obtain $\widetilde{\bm{x}}_2$, and subsequently subtract this result from the received signal as:
\begin{equation}
    \bm y = \bm y - h_2\widetilde{\bm x}_2.
\end{equation}
We further estimate $\bm x_3$ as:
\begin{equation}
    \widehat{\bm x}_3 = h_3^{-1}\bm{y}.
\end{equation}
The semantic decoding of $\widehat{\bm x}_3$ into $\widehat{T}_3$ can then be carried out immediately. The semantic decoding of other texts can be done in a similar manner, as illustrated in Fig.~\ref{fig:semantic_decoder}.

The above method can be thought of as a semantic generalization of the traditional SIC scheme. Nevertheless, it does not take into account the semantic correlation between the different texts. In particular, if $T_i$ and $T_w$ are on the same topic (e.g., they are two reports of the same news), then knowing $T_i$ already can significantly help the decoding of $T_w$. Thus, we propose to improve the semantic decoder by incorporating the previously recovered texts as side information. Specifically, when decoding $T_i$, we use the updated received signal $\widehat{\bm x}_i$, from which the signals of the previous texts $T_1,\ldots,T_{i-1}$ have been removed, and $(\widetilde{\bm r}^1_j,\widetilde{\bm r}^2_j,\ldots,\widetilde{\bm r}^{i-1}_j)$ as inputs of integrated feature generator as shown in Fig.~\ref{fig:deepsc_si_combiner3} to generate $\bm g$:
\begin{equation}
    \bm g^i_{j} = \mathcal{IFG}_i(\widehat{\bm r}^{i}_j,\widetilde{\bm r}^1_j,\widetilde{\bm r}^2_j,\ldots,\widetilde{\bm r}^{i-1}_j),
\end{equation}
where $\bm g^i_{j}$ is the new feature we generate by $\mathcal{IFG}_i$ to facilitate decoding the $j$th sentence of user terminal $i$ and the integrated feature generator component assigned to user terminal $i$ is defined to be:
\begin{align}
&\mathcal{IFG}_i(\widehat{\bm r}^{i}_j,\widetilde{\bm r}^1_j,\widetilde{\bm r}^2_j,\ldots,\widetilde{\bm r}^{i-1}_j)  = \notag\\ 
&\quad\text{ReLU}(\bm{\theta}_i((\bm{\pi}_i(\widehat{\bm r}_{i}^j),\bm{\Omega}_i^{1}(\widetilde{\bm r}_{1}^j), \dots,\bm{\Omega}_i^{i-1}(\widetilde{\bm r}_{i-1}^j)) )+ \widehat{\bm r}_{i}^j),
\label{ifg}
\end{align}
with the functions $\Omega_i^w(\cdot)$, $\bm\theta_i(\cdot)$, and $\pi_i(\cdot)$ defined in Fig.~\ref{fig:deepsc_si_combiner3}, while ReLU denoting the rectified linear unit. Subsequently, $\widehat{\bm u}^i_{j,\ell}$ can be obtained as:
\begin{equation}
    \widehat{\bm u}^i_{j,\ell} =  \mathrm{AD}_i(\bm g^i_{j,\ell}).
\end{equation}
After obtaining \(\widehat{\bm u}^i_{j,\ell}\), the decoding proceeds in the same way as the decoding of \(\widehat{\bm u}^1_{j,\ell}\) in \ref{subsec:reverse}, ultimately leading to \(T_i\).

\section{Neural Network Training}
\label{sec:training}

\subsection{Loss Function Design}
\label{loss_design}

We start by describing the primary objective of our system: semantically accurate sentence transmission from each user $i$ to the BS. For user $i$, this is accomplished by minimizing the cross-entropy between the source text $T_i$ and the reconstructed sentence $\widehat{T_i}$. The expression for cross-entropy loss is given by:
\begin{multline}
\label{loss_ce}
L_\texttt{CE}(T_{i},\widehat{T}_i) = -\frac{1}{L_i}\sum_{j=1}^{L_i}\sum_{(n_j,w_j)} q_{n_j}(V_{w_j})  \log_2p_{n_j}(V_{w_j})\\
- \frac{1}{L_i}\sum_{j=1}^{L_i}\sum_{(n_j,w_j)} (1-q_{n_j}(V_{w_j})) \log_2(1-p_{n_j}(V_{w_j})),
\end{multline}
where, in $j$th sentence, $q_{n_j}(V_{w_j})\in\{0,1\}$ is the ground-truth label such that $q_{n_j}(V_{w_j})=1$ if $n_j$th word is correctly identified as $V_{w_j}$ and $q_{n_j}(V_{w_j})=0$ otherwise, while $p_{n_j}(V_{w_j})\in[0,1]$ is the soft decision that reflects the likelihood of the $n$th word being $V_{w_j}$.

\subsection{Initial Training for $K$ Users in the MAC}
\label{trainingSP}

\begin{algorithm}[t]
\caption{Initial Training for a $K$-User MAC}
\label{alg:train_K}
\begin{algorithmic}[1]
    \State \textbf{Input:} The knowledge sets $\mathcal{K}_1$, \dots, $\mathcal{K}_{K}$
    \For{user $i\in\{1,2,\ldots,K\}$}
      \State  \(T_i \leftarrow\) \textsf{BatchSource}($\mathcal{K}_i$)
      \State \((x^i_1,x^i_2,\ldots,x^i_M)^\top \leftarrow \textsf{ENC}_i(T_i)  \)
    \EndFor
\State Transmit $(x^i_1,x^i_2,\ldots,x^i_M)^\top$ on the MAC channel

\State Receive signal \( \bm y \)
\State $\{\widehat{T}_1,\widehat{T}_2,\ldots,\widehat{T}_K\} \leftarrow \textsf{DEC}(\bm y)$
\State  Compute loss function \( L_\texttt{joint} \) by \eqref{joint}
\State Train $\textsf{ENC}_1$, \dots, $\textsf{ENC}_{K}$, $\textsf{DEC}$ by backpropagation for the loss $L_\texttt{joint}$
\State \textbf{Output:} $\textsf{ENC}_1$, \dots, $\textsf{ENC}_{K}$, $\textsf{DEC}$

\end{algorithmic}
\end{algorithm}

We first consider how to perform training given $K$ users; the next subsection discusses how to adjust the initially trained neural network when new users are added to the MAC.

For the fixed $K$ users, we optimize the parameters of the neural network via the Adam optimizer \cite{kingma2014adam}, for the loss function:
\begin{equation}
\label{joint}
L_\texttt{joint} = \sum_{i=1}^K L_\texttt{CE}(T_{i},\widehat{T}_i).
\end{equation}

We decide the starting point by the pretraining method. Specifically, we train the neural network with respect to each single user separately in an end-to-end fashion, without considering the inter-user interference, thus quickly obtaining the encoder and decoder for each link. We will show in Section \ref{sec:experiments} that the above starting point yields faster convergence of the loss function during the training session. The resulting algorithm is summarized in Algorithm \ref{alg:train_K}.

\begin{algorithm}[t]
\caption{Partial Retraining After $n$ New Users Join MAC}
\label{alg:train_partial}
\begin{algorithmic}[1]
\State \textbf{Initialization:} Execute Algorithm \ref{alg:train_K}
    \State \textbf{Input:} The knowledge sets $\mathcal{K}_1$, \dots, $\mathcal{K}_{K+n}$
    \For{user $i\in\{1,2,\ldots,K+n\}$}
      \State  \(T_i \leftarrow\) \textsf{BatchSource}($\mathcal{K}_i$)
      \State \((x^i_1,x^i_2,\ldots,x^i_M)^\top \leftarrow \textsf{ENC}_i(T_i)  \)
    \EndFor
\State Transmit $(x^i_1,x^i_2,\ldots,x^i_M)^\top$ on the MAC channel

\State Receive signal \( \bm y \)
\State $\{\widehat{T}_1,\widehat{T}_2,\ldots,\widehat{T}_K\} \leftarrow \textsf{DEC}(\bm y)$
\State  Compute loss function \( L_\texttt{FP}(K+1,K+n) \) by \eqref{loss_full}
\State Train $\textsf{ENC}_{K+1}$, \dots, $\textsf{ENC}_{K+n}$, $\textsf{DEC}$ by backpropagation for the loss $ L_\texttt{FP}(K+1,K+n)$

\State \textbf{Output:} $\textsf{ENC}_{K+1}$, \dots, $\textsf{ENC}_{K+n}$, $\textsf{DEC}$
\end{algorithmic}
\end{algorithm}

\begin{figure}[t]
\begin{minipage}[b]{8cm}
    \centering
\includegraphics[width=0.7\linewidth]{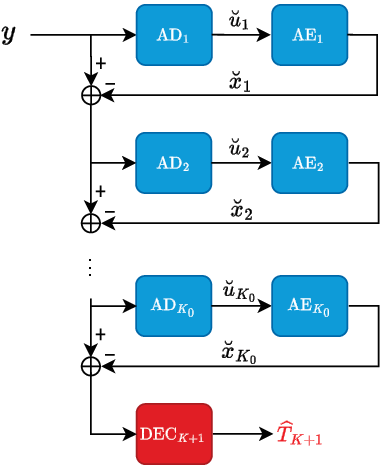}
\caption{Phase I of the adapted algorithm for one-user addition: remove from $y$ the signals of those old users who are stronger than the new user $K+1$, and then recover the message $T_{K+1}$. It is important to note that $\breve{u}_i$ does not serve to recover text; its primary role is to generate $\breve{x}_i$, which assists in the generation of $T_{K+1}$.}

\label{fig:CPPR_part1}
\end{minipage}
\end{figure}

\subsection{Retraining After $n$ New Users Join MAC}

What if some new users are added to the existing $K$-user MAC after the neural networks have been well-trained according to Algorithm \ref{alg:train_K}, i.e., how do we perform retraining? We propose two possible methods: the full retraining and the partial retraining. The full retraining method is to configure the neural network from scratch. We take the output of Algorithm \ref{alg:train_K} as the starting point for the $K$ existing users, and initialize the newly added $n$ users by the pretraining method as discussed below \eqref{joint}. We then retrain the neural network by using Algorithm \ref{alg:train_K} for the following loss function:
\begin{equation}
\label{loss_full}
L_\texttt{FP}(\xi,\rho) = \sum^{\rho}_{i=\xi} \tau_i L_\texttt{CE}(T_{i},\widehat{T}_i),
\end{equation}
where $\xi>0$ is the smallest index of trained users, $\rho>0$ is the number of trained users and $\tau_i>0$ is a tradeoff factor between the old users and the new users. 

However, full retraining of the whole neural network as stated above is computationally formidable. The second method, referred to as the partial retraining, only trains the neural networks for the newly added $n$ users, while preserving those for the existing $K$ users, which is illustrated in Algorithm \ref{alg:train_partial}. The key idea of this method is: when the messages of the new $n$ users can be decoded successfully, the existing $K$ users can be dealt with as if the new users did not exist, and thereby the previously trained neural network for the $K$ users can be reused. Thus, during decoding, the second method comprises two stages: (i) decode the messages for the new $n$ users in the presence of the interference from the existing $K$ users; (ii) decode the messages for the existing $K$ users as if the new $n$ users did not exist. 

\begin{figure}[t]
\begin{minipage}[b]{8cm}
    \centering
\includegraphics[width=0.9\linewidth]{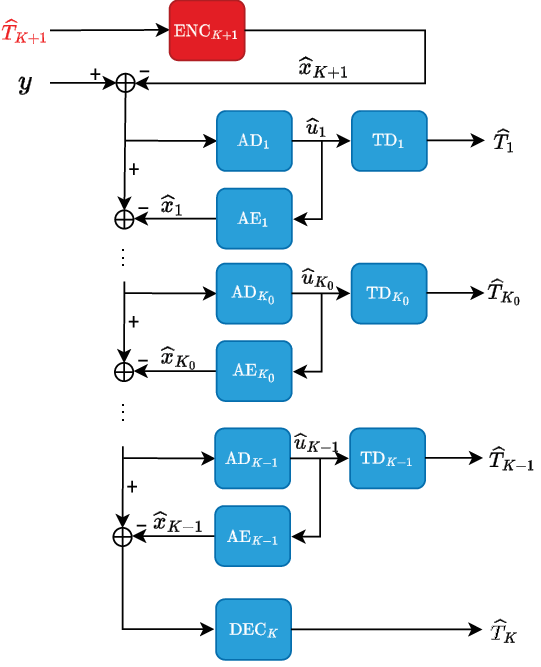}
\caption{Phase II of the adapted algorithm for one-user addition: remove from $y$ the signal of the new user $K+1$, and then decode the messages of the old users as if the new user does not exist.}
\label{fig:CPPR_part2}
\end{minipage}
\end{figure}

For instance, assume that there are a total of $K$ users in the multiple access network, and assume without loss of generality that satisfies \eqref{sic_seq}. Now a new user indexed $K+1$ joins the multiple access channel. Assume that
\begin{multline}
\label{add_new}
P_1|h_1|^2 \ge \ldots \ge P_{K_0}|h_{K_0}|^2\ge P_{K+1}|h_{K+1}|^2 \\ \ge P_{K_0+1}|h_{K_0+1}|^2\ge \ldots\ge P_{K}|h_{K}|^2.
\end{multline}
Thus, the old users $\{1,2,\ldots,K\}$ can be partitioned into two groups $G_1=\{1,2,\ldots,K_0\}$ and $G_2=\{K_0+1,K_0+2,\ldots,K\}$. We then train the encoder and decoder for the new user $K+1$ as before, without modifying those for the old users.

After training for the new user alone, the encoder part works in the same way as before, but the decoder part is changed. The new decoder consists of two phases. In phase I, we try decoding the messages in group $G_1$ and then remove their signals from $y$ to facilitate the decoding for the new user. Thus, the message of the new user, $T_{K+1}$, is recovered at the end of phase I. Next, in phase II, we begin by removing the signal of $T_{K+1}$ from the received signal $y$. Subsequently, the messages from the $K$ old users can be decoded as if the new user does not exist. This is the rationale for adapting the semantic communication method for the new user without retraining the encoders and decoders for the existing users.

We remark that the above two coding processes lead to distinct decoders, as summarized in Algorithm~\ref{alg:decoding} and Algorithm~\ref{alg:reverse_sic}.

\begin{algorithm}[t]
\caption{Coding Process After Full Retraining}
\label{alg:decoding}
\begin{algorithmic}[1]
    \State \textbf{Input:} The knowledge sets $\mathcal{K}_1$, \dots, $\mathcal{K}_{K+n}$
    \For{user $i\in\{1, 2, \dots, K+n\}$}
      \State  \(T_i \leftarrow\) \textsf{BatchSource}($\mathcal{K}_i$)
      \State \((x^i_1,x^i_2,\ldots,x^i_M)^\top \leftarrow \textsf{ENC}_i(T_i)  \)
    \EndFor
\State Transmit $(x^i_1,x^i_2,\ldots,x^i_M)^\top$ on the MAC channel

\State Receive signal \( \bm y\)
\State $\{\widehat{T}_{1},\dots,\widehat{T}_{K},\widehat{T}_{K+1},\ldots,\widehat{T}_{K+n}\} \leftarrow \textsf{DEC}(\bm y)$

\State \textbf{Output:} The estimated sentences $\widehat{T}_1$, $\widehat{T}_2$, \dots, $\widehat{T}_{K+n}$
\end{algorithmic}
\end{algorithm}

\begin{algorithm}[t]
\caption{Coding Process After Partial Retraining}
\label{alg:reverse_sic}
\begin{algorithmic}[1]
\State \textbf{Input:} The knowledge set $\mathcal{K}_1$, \dots, $\mathcal{K}_{K+n}$
\State Decode the signals of the old users $\{1,\ldots,K_0\}$ and the new users $\{K+1,\ldots,K+n\}$ by the proposed semantic SIC to obtain $\{\widehat{T}_{K+1}, \dots, \widehat{T}_{K+n}\}$
\State $\bm y \leftarrow \bm y-\sum^{K+n}_{i=K+1} h_i\textsf{ENC}_{i}(\widehat{T}_{i})$ 
\State $\{\widehat{T}_{1},\ldots,\widehat{T}_{K}\} \leftarrow \textsf{DEC}(\bm y)$

\State \textbf{Output:} The estimated sentences $\widehat{T}_1$, $\widehat{T}_2$, \dots, $\widehat{T}_{K+n}$
\end{algorithmic}
\end{algorithm}

\begin{figure}[t]
\begin{minipage}[b]{8cm}
\includegraphics[width=1.0\linewidth]{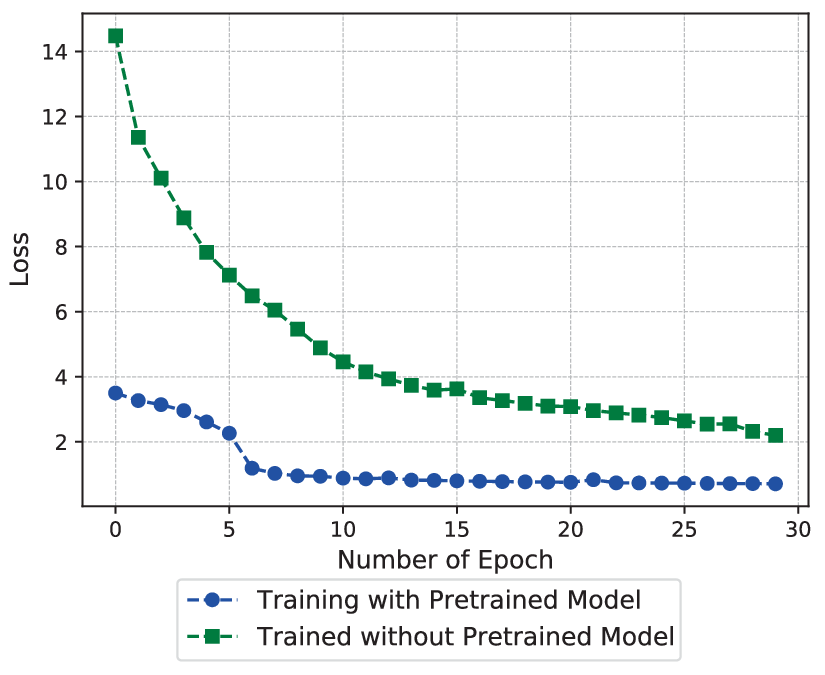}
\caption{Loss Comparison between the pretraining scheme and the retraining scheme when there are 5 users in the MAC.}
\label{fig:loss_5users}
\end{minipage}
\end{figure}

\begin{figure*}[ht]
\begin{minipage}{1\textwidth}
    \centering
\includegraphics[width=11cm]{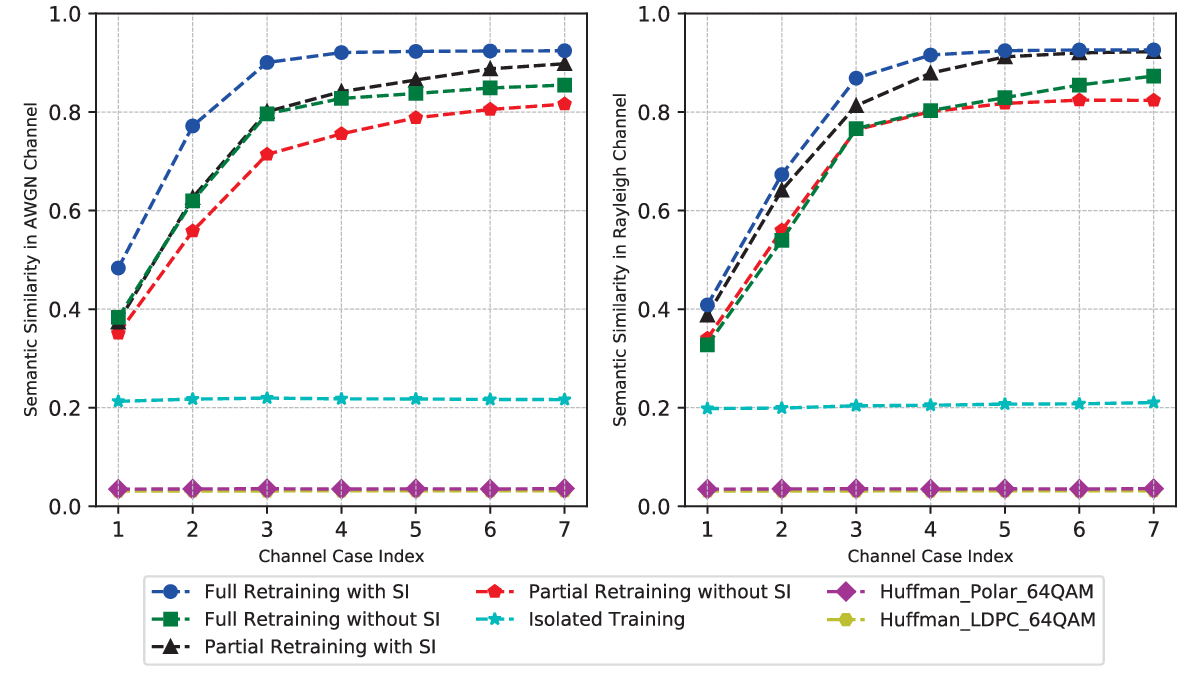}    
\end{minipage}
\caption{Minimal semantic similarity across 2+1 users in the AWGN case (left) and the Rayleigh fading case (right) under different channel cases in Table \ref{tab:SNRs}.}
\label{fig:simi_3users}
\end{figure*}

\begin{figure*}[t]
\centering
\includegraphics[width=11cm]{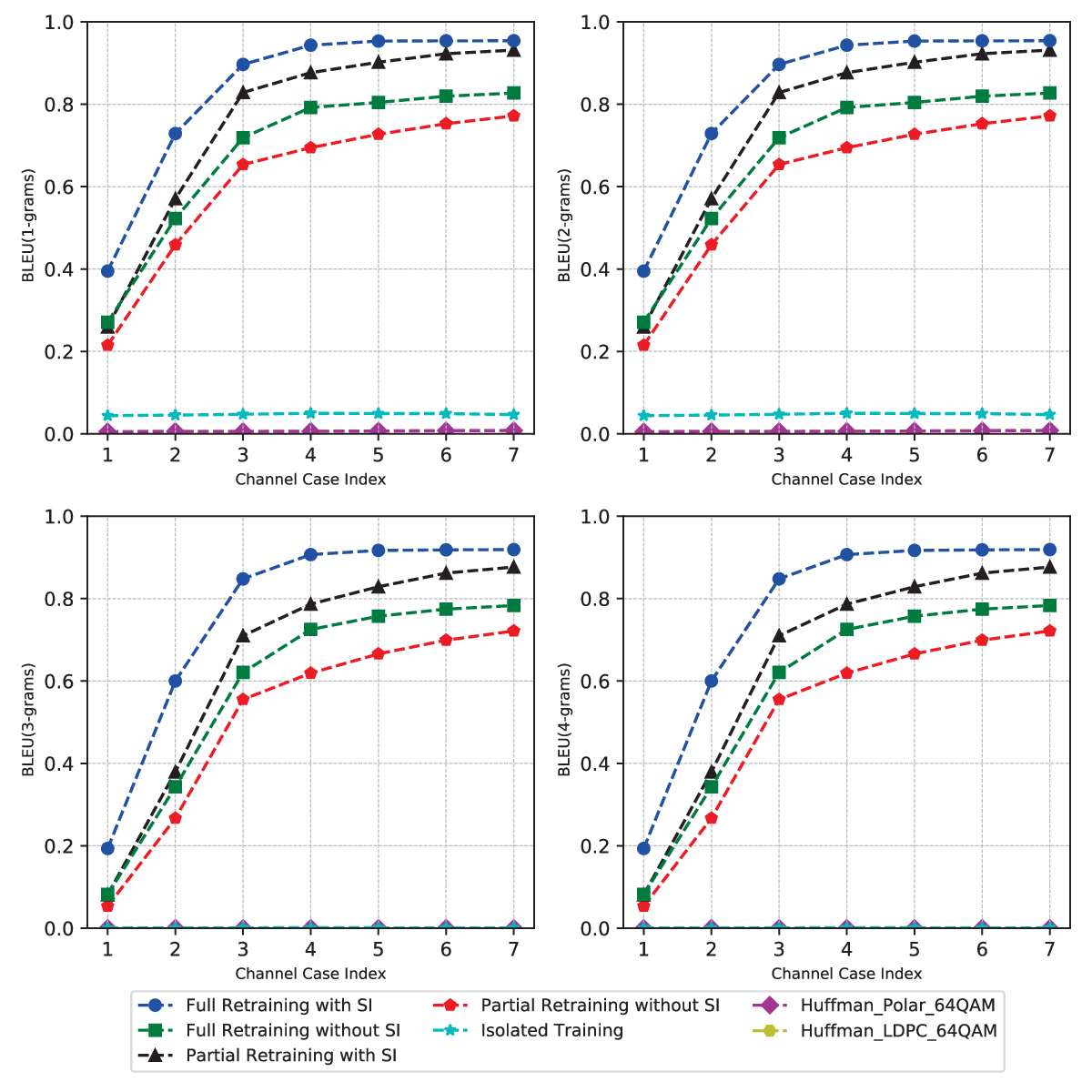}
\caption{Minimal BLEU score across 2+1 users in the AWGN case under different channel cases in Table \ref{tab:SNRs}.}
\label{fig:bleu3_awgn}
\end{figure*}

\section{Numerical Experiments}
\label{sec:experiments}
\subsection{Setup}
The dataset utilized in this study is the Stanford Natural Language Inference (SNLI) Corpus \cite{bowman2015large}, which comprises approximately 550,000 pairs of English sentences. We select the dataset SNLI Corpus because it exhibits much stronger inter-sentential connections than other existing datasets (e.g., the
Europarl in \cite{koehn-2005-europarl}), as demonstrated in the literature \cite{lee2023seq2seq,fradelos2023using} on natural language processing. Specifically, for each sentence, the SNLI Corpus dataset provides three classes of reference sentences---which respectively correspond to entailment, contradiction, and neutrality. For instance, the following sentence:
$$``\small\textsf{A man in an orange vest is leaning over a pickup truck.}"\\
$$
is paired to an entailment:
$$``\small\textsf{A man is touching a truck.}"\\
$$
in the SNLI Corpus. Clearly, the extensive inter-sentential connections facilitate learning the impact of context on words, thereby improving the semantic feature extraction. The dataset is preprocessed to ensure sentence lengths range from $4$ to $20$ words and is then divided into training and test sets.

All results in this paper discuss the scenario of adding users to a system. Firstly, in a single-user scenario, each user is trained for 50 epochs. Secondly, for a given number of $K$ users, the networks are trained for an additional 30 epochs. Lastly, when $n$ new users join in the $K$-user system, the network is retrained either fully or partially for 30 epochs. We employ the Adam optimizer \cite{kingma2014adam} with a learning rate of $1 \times 10^{-3}$, $\beta = (0.9, 0.98)$, $\epsilon = 1 \times 10^{-8}$, and a weight decay of $5 \times 10^{-4}$. Moreover, a greedy decoder with repetition penalty \cite{melas2018training} is utilized during the testing stage. Additionally, in this experiment, we allocate 18 symbols for each word. We compare the performance of our proposed methods against established benchmark methods. Additionally, we have taken into account two traditional baseline scenarios that employ conventional source and channel coding techniques:
\begin{itemize}
    \item Huffman coding \cite{huffman1952method} combined with LDPC coding \cite{ryan2004introduction} using $64$-QAM modulation.
    \item Huffman coding \cite{huffman1952method} paired with Polar coding \cite{bioglio2020design} utilizing $64$-QAM modulation.
\end{itemize}
In contrast, five deep learning-related methods including our proposed methods are considered with the settings below:
\begin{itemize}
        \item Full retraining with SI: proposed method, in which the models are fully retrained with side information.
        \item Full retraining without SI: benchmark method, in which the models are fully retrained without side information.
        \item Partial retraining with SI: proposed method, in which the models are partially retrained with side information.
        \item Partial retraining without SI: benchmark method, in which the models are partially retrained without side information.
        \item Isolated training without SI: benchmark method, in which the models are trained in a single-user fashion without side information.
\end{itemize}

In addition to adopting the sentence similarity defined in \eqref{simi} as a metric, we also measure the semantic error using the Bilingual Evaluation Understudy (BLEU) score \cite{bleu} to compare the original sentence and the recovered sentence. A higher BLEU score indicates better transmission performance. The BLEU score for the original text $T_i$ and the recovered text $\widehat{T}_i$ is calculated as
\begin{equation}
\hspace{-0.5cm}
\text{BLEU}(T_i,\widehat{T}_i) = \frac{1}{L_i}\sum^{L_i}_{j=1}\exp \bigg(\text{BP} + \sum_{n_j} u_{n_j} \log p_{n_j}\bigg),
\end{equation}
where $\text{BP}$ is defined as
\begin{equation}
\text{BP} =\min \bigg(1 - \frac{\mathrm{len}({\widehat{S}_j})}{\mathrm{len}({S_j})}, 0\bigg).
\end{equation}
In this formula, $u_{n_j}$ represents the weights assigned to $n$-grams and $p_{n_j}$ is the precision of $n$-grams, which is defined as
\begin{equation}
p_{n_j} = \frac{\sum_{k} \min (C_k(\widehat{S}_j), C_k(S_j))}{\sum_{k} \min (C_k(S_j))},
\end{equation}
where $C_k(\cdot)$ denotes the count of the $k$th element in $n$-grams.

\subsection{Simulation Results}

Fig.~\ref{fig:loss_5users} compares the training loss versus epochs for two cases: with and without the pretrained model. The blue line represents \textit{Training with Pretrained Model}, which rapidly decreases and converges at a low level after around five epochs. The green line represents \textit{Training without Pretrained Model}, which starts with a higher loss and decreases more slowly, and ends up with a higher loss than the pretrained approach. This indicates that employing the pretrained model not only accelerates training but also leads to lower loss.

\begin{table}
    \centering
    \caption{Cases in the 2+1 users experiments (in dB)}
    \begin{tabular}{c||c|c|c}
        \hline
         Case Index& User 1 & User 2 & User 3\\
         \hline\hline
         1 & -3.29  & -5.95 & -9 \\
         2 & -0.29  & -2.95 & -6 \\
         3 & 2.71 & 0.05 & -3 \\
         4 & 5.71 & 3.05 & 0 \\
         5 & 8.71 & 6.05 & 3 \\
         6 & 11.71 & 9.05 & 6 \\
         7 & 14.71 & 12.05 & 9 \\
         \hline
    \end{tabular}
    \label{tab:SNRs}
\end{table}

\begin{table}
    \centering
    \caption{Time of Full Retraining and Partial Retraining while adding one user to a two-user MAC}
    \begin{tabular}{c||c|c}
        \hline
         & Full Retraining  & Partial Retraining \\
         \hline
         With SI & 2.09s/it &  1.43s/it\\
         \hline
         Without SI & 1.62s/it &  0.99s/it\\
         \hline
    \end{tabular}
    \label{tab:time}
\end{table}

First, we examine the impact of adding one user to a two-user MAC (namely, the ``2+1'' case) for seven distinct channel conditions as shown in Table \ref{tab:SNRs}. Fig.~\ref{fig:simi_3users} illustrates the minimal semantic similarity for 2+1 users in both AWGN (left) and Rayleigh fading (right) scenarios. In the AWGN case, the \textit{Full Retraining with SI} method achieves superior performance by achieving higher semantic similarity across all seven channel cases. It has considerable advantages over those traditional approaches based on the 64-QAM modulation. Moreover, \textit{Partial Retraining with SI} exhibits comparable performance with the \textit{Full Retraining without SI} in the first three channel cases, thus implying that side information is useful in further enhancing the efficiency of partial retraining. However, in the latter four channel cases, \textit{Partial Retraining with SI} surpasses \textit{Full Retraining without SI}, indicating that side information is beneficial even in the training resource-limited situation. In the Rayleigh fading scenario, \textit{Partial Retraining with SI} consistently outperforms \textit{Full Retraining without SI}, reinforcing the advantage of incorporating side information. This trend suggests that side information is particularly useful in maintaining high semantic similarity in harsh wireless environments. In contrast, the traditional methods  \textit{Huffman\_Polar\_64QAM} and \textit{Huffman\_LDPC\_64QAM} struggle to accomplish the transmission tasks in both scenarios. For these two methods, although they reach high SNR, the SINR levels are fairly low due to the poor interference management. This reveals the limitations of traditional methods in adapting to dynamic channel conditions, and also highlights the important role played by side information in improving
the overall network performance from a semantic point of view.
\begin{table*}
    \centering
    \caption{Cases in the 3+2 users experiments (in dB)}
    \begin{tabular}{c||c|c|c|c|c}
        \hline
         Case Index& User 1 & User 2 & User 3& User 4 & User 5\\
         \hline\hline
         1 & -3.29  & -5.95& -6.27&-7.49 & -9 \\
         2 & -0.29  & -2.95&-3.27 & -4.49& -6 \\
         3 & 2.71 & 0.05 &-0.27 & -1.49& -3 \\
         4 & 5.71 & 3.05 & 2.73 & 1.51 & 0 \\
         5 & 8.71 & 6.05 &5.73 &4.51 & 3 \\
         6 & 11.71 & 9.05&8.73 &7.51 & 6 \\
         7 & 14.71 & 12.05&11.73 &10.51 & 9 \\
         \hline
    \end{tabular}
    \label{tab:SNRs_5}
\end{table*}

\begin{figure*}[t]
\centering
\includegraphics[width=11cm]{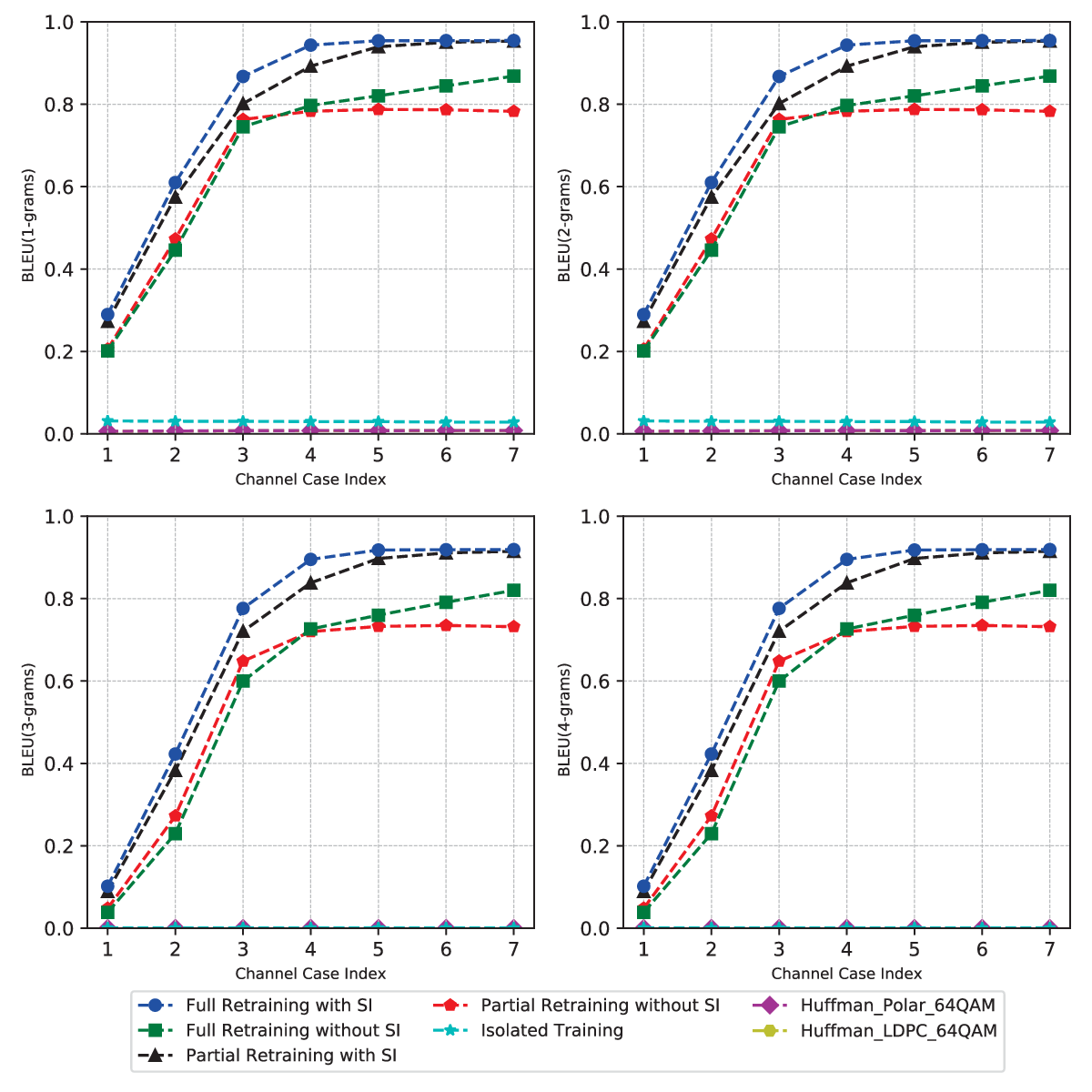}
\caption{The minimal BLEU score for 2+1 users in the Rayleigh fading case under different channel cases in Table \ref{tab:SNRs}.}
\label{fig:bleu3_rayl}
\end{figure*}

Fig.~\ref{fig:bleu3_awgn} compares the minimal BLEU scores for 2+1 user transmissions (i.e., a new user is added to an existing 2-user MAC) in the AWGN channel across various channel conditions as summarized in Table II. The results are categorized by the n-gram order, with each subplot representing BLEU scores for 1-grams through 4-grams. The \textit{Full Retraining with SI} method achieves the highest BLEU scores in all the subplots. \textit{Partial Retraining with SI} gives significant improvements upon \textit{Full Retraining without SI}, especially for a higher n-gram orders, demonstrating the value of side information in transmission tasks. In contrast, \textit{Isolated Training} and traditional methods like \textit{Huffman\_Polar\_64QAM} and \textit{Huffman\_LDPC\_64QAM} obtain the lowest BLEU scores, due to their limited capability in managing the complexities of AWGN channels. The above results again show that the side information is quite helpful for the gain it brings regarding the BLEU score.

Fig.~\ref{fig:bleu3_rayl} illustrates the BLEU scores for three-user transmissions in a Rayleigh fading channel under various channel conditions in Table II. The \textit{Full Retraining with SI} method excels among all the methods, achieving the highest BLEU scores. The \textit{Partial Retraining with SI} method also performs well, being close to the full retraining method, especially in the first three channel cases. However, as channel conditions improve, the \textit{Full Retraining with SI} method starts to outperform. In contrast, the traditional methods, \textit{Huffman\_Polar\_64QAM} and \textit{Huffman\_LDPC\_64QAM}, acheive much lower BLEU scores.

\begin{figure}[t]
\begin{minipage}[b]{8cm}
\includegraphics[width=8.2cm]{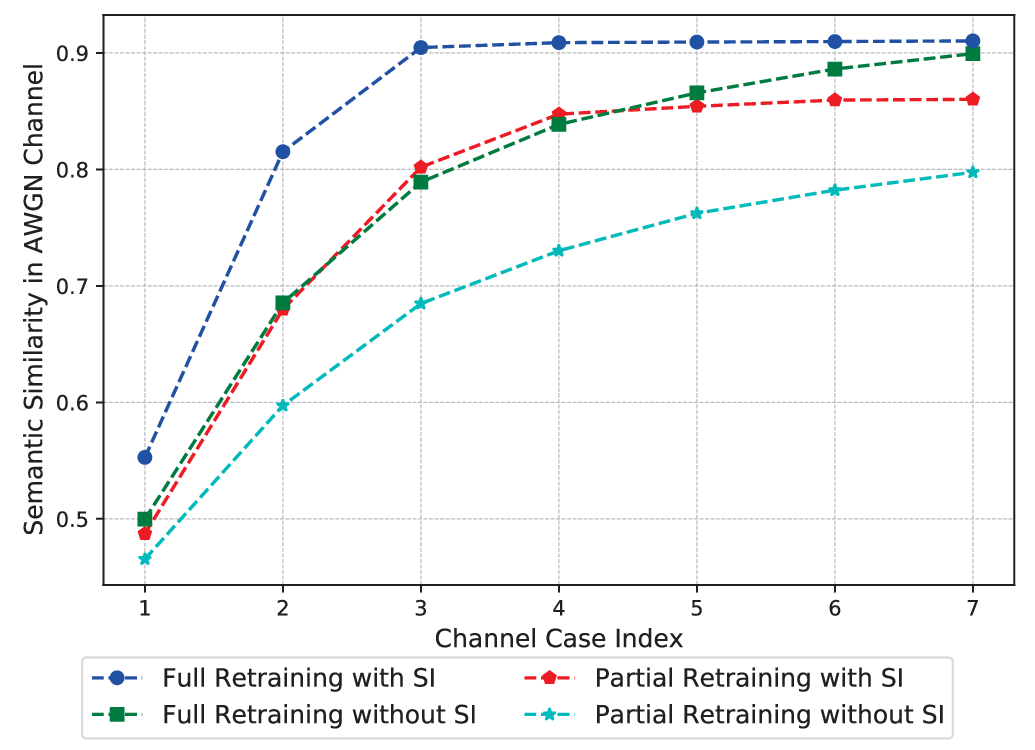}
\caption{The minimal semantic similarity for 3+2 users of in the AWGN case under different channel cases in Table \ref{tab:SNRs_5}.}
\label{fig:5users_AWGN}
\end{minipage}
\end{figure}

Fig.~\ref{fig:5users_AWGN} illustrates that \textit{Full Retraining with SI} achieves the highest semantic similarity for 3+2 users (i.e., 2 new users are added to an existing 3-user MAC) in the AWGN channel. \textit{Partial Retraining with SI} attains similar performance. The above two methods outperform the methods without SI.

Table~\ref{tab:time} provides a comparison of training times for full and partial retraining with SI versus that without SI when a new user is added to an existing two-user MAC. \textit{Full Retraining with SI}, despite having the longest training time of 2.09 seconds per iteration, ends up with the best performance. In contrast, \textit{Partial Retraining without SI} is the most computation-efficient method, completing each iteration in merely 0.99 seconds, but this efficiency comes at the cost of potentially large performance loss. Notably, \textit{Partial Retraining with SI} not only achieves desirable performance but also cuts down training time by over 30\% as compared to \textit{Full Retraining with SI}, making itself a balanced choice between efficiency and performance.

\section{Conclusion}
\label{sec:conclusion}
This paper considers the extension of the DeepSC scheme to the MAC scenario, which significantly enhances how semantic processing is integrated within communication systems to handle inter-link interference. By implementing SIC in the semantic domain and leveraging semantic correlations for decoding, this approach innovatively mitigates interference and improves decoding accuracy. The shift from the full retraining to the partial retraining addresses the scalability challenges posed by the addition of new users, offering a flexible and efficient adaptation. For the future research work, we would like to develop the proposed semantic communication method further in two respects. First, we aim at a MIMO extension; in fact, some existing works \cite{xie2022task,liang2023semantic} already consider the MIMO semantic communication yet without SIC. Second, we aim at a multi-modal extension that accounts for other messages (such as images and audio); it requires verifying that the proposed scheme can preserve semantic similarity under other metrics aside from BERT and BLEU for text.

\label{sec:refs}
\bibliographystyle{IEEEtran}
\bibliography{IEEEabrv,refs}

\begin{IEEEbiographynophoto}{Mingxiao Li}
(Graduate Student Member, IEEE) received the B.E. degree from Beijing University of Posts and Telecommunications, Beijing, China, in 2022. He is currently pursuing his Ph.D. degree with The Chinese University of Hong Kong (Shenzhen). His research interests include semantic communications, wireless communications and machine learning.
\end{IEEEbiographynophoto}

\begin{IEEEbiographynophoto}{Kaiming Shen}
(Senior Member, IEEE) received the B.Eng. degree in information security and the B.Sc. degree in mathematics from Shanghai Jiao Tong University, China in 2011, and then the M.A.Sc. degree in electrical and computer engineering from the University of Toronto, Canada in 2013. After working at a tech startup in Ottawa for one year, he returned to the University of Toronto and received the Ph.D. degree in electrical and computer engineering in early 2020. Dr. Shen has been with the School of Science and Engineering at The Chinese University of Hong Kong (CUHK), Shenzhen, China as a tenure-track assistant professor since 2020. His research interests include optimization, wireless communications, information theory, and machine learning.
Dr. Shen received the IEEE Signal Processing Society Young Author Best Paper Award in 2021, the CUHK Teaching Achievement Award in 2023, and the Frontiers of Science Award at the International Congress of Basic Science in 2024. Dr. Shen currently serves as an Editor for IEEE Transactions on Wireless Communications.
\end{IEEEbiographynophoto}

\begin{IEEEbiographynophoto}{Shuguang Cui}
(Fellow, IEEE) received the Ph.D. degree in Electrical Engineering from Stanford University, California, USA, in 2005. Afterwards, he has been working as an Assistant, Associate, Full, Chair Professor in Electrical and Computer Engineering at The University of Arizona, Texas A\&M University, UC Davis, and The Chinese University of Hong Kong at Shenzhen respectively. He has also served as the Executive Dean for the School of Science and Engineering at The Chinese University of Hong Kong, Shenzhen, China, the Executive Vice Director at Shenzhen Research Institute of Big Data, and the Director for Shenzhen Future Network of Intelligence Institute (FNii Shenzhen), Shenzhen, China. His current research interests focus on the merging between AI and communication neworks. He was selected as the Thomson Reuters Highly Cited Researcher and listed in the Worlds’ Most Influential Scientific Minds by ScienceWatch in 2014. He was the recipient of the IEEE Signal Processing Society 2012 Best Paper Award. He has served as the General Co-Chair and TPC Co-Chairs for many IEEE conferences. He has also been serving as the Area Editor for IEEE Signal Processing Magazine, and Associate Editors for IEEE Transactions on Big Data, IEEE Transactions on Signal Processing, IEEE JSAC Series on Green Communications and Networking, and IEEE Transactions on Wireless Communications. He has been
the Elected Member for IEEE Signal Processing Society SPCOM Technical Committee (2009–2014) and the Elected Chair for IEEE ComSoc Wireless Technical Committee (2017–2018). He is a member of the Steering Committee for IEEE Transactions on Big Data and the Chair of the Steering Committee for IEEE Transactions on Cognitive Communications and Networking. He is also the Vice Chair of the IEEE VT Fellow Evaluation Committee and a member of the IEEE ComSoc Award Committee. He was elected as an IEEE Fellow in 2013, an IEEE ComSoc Distinguished Lecturer in 2014, and IEEE VT Society Distinguished Lecturer in 2019. In 2020, he won the IEEE ICC Best Paper Award, ICIP Best Paper Finalist, the IEEE Globecom Best Paper
Award. In 2021, he won the IEEE WCNC Best Paper Award. In 2023, he won
the IEEE Marconi Best Paper Award, got elected as a Fellow of both Canadian Academy of Engineering and the Royal Society of Canada, and starts to serve as the Editor-in-Chief for IEEE Transactions on Mobile Computing.
\end{IEEEbiographynophoto}

\end{document}